\documentstyle[12pt]{article}

\font\ss=msbm10

\newcommand{\be}{\begin{equation}}
\newcommand{\ee}{\end{equation}}
\newcommand{\bean}{\begin{eqnarray*}}
\newcommand{\eean}{\end{eqnarray*}}
\newcommand{\bea}{\begin{eqnarray}}
\newcommand{\eea}{\end{eqnarray}}
\newcommand{\RR}{\hbox{\ss R}}
\newcommand{\diag}{{\rm diag}}
\newcommand{\Tr}{{\rm Tr}}
\newcommand{\sq}{\sqrt{|g|}}
\newcommand{\sg}{\sqrt{|\gamma|}}

\author{M. G. Ivanov\thanks{e-mail: mgi@mi.ras.ru}}
\title{Delocalized membrane model}
\date{July 13, 2001}

\begin{document}

\maketitle

\abstract{
  A model considered in the paper generalizes membrane theory
 to the case of delocalized membranes.
  The model admits covariant formulation, which involves
 no constraints.
  It generalizes the notion of membrane to the case of 
 smooth distribution of non-intersecting membranes.
  A generalization of p-brane solution with delocalized membranes
 is presented.
}

\section{Introduction}

  Membrane theories \cite{GSW,BN,VV,Duff} consider membranes as mapping $x$
 from manifold ${\bf V}$ of dimension $n$ to (pseudo)Riemannian manifold
 ${\bf M}$ of dimension $D>n$
 (``$({\bf V\to M})$-approach'', or ``$({\bf V\to M})$-theories'').
  The manifold ${\bf M}$ is interpreted as space-time (``bulk'').
  The image $x({\bf V})$ is interpreted as membrane.
  Under this formulation the action of membrane theory includes terms
 of two types, one of which involves integration over ${\bf V}$,
 the other involves integration over ${\bf M}$.
  Similarly the range of equation of motion is either ${\bf V}$ or ${\bf M}$.
  Membrane appears in bulk equation in the form of singular sources localized
 at $x({\bf V})$.
  The goal of the present paper is to provide a covariant method of
 regularization of closed membrane theories.

  To regularize these singularities one has to replace fields defined at
 ${\bf V}$ by some bulk fields.
  The procedure of regularization can be considered as replacement of
 common membrane, which has zero thickness by some sort of thick membrane,
 which corresponds to continuous distribution of infinitely light membranes.

  The approach presented in the paper
 (``$({\bf M\to F})$-approach'')
 replaces mapping $x:{\bf V\to M}$
 by mapping $\varphi:{\bf M\to F}$, where ${\bf F}$ is a manifold of dimension
 $D-n$.
  If for some point $\phi\in{\bf F}$ inverse image $\varphi^{-1}(\phi)$
 is submanifold of dimension $n$, then it has to be considered
 as infinitely light membrane.
  The generalization of p-brane solution with delocalized membranes
 is presented in section \ref{p-b-sec} as an example of
 $({\bf M\to F})$-approach.

  In comparison with $({\bf V\to M})$-theories in $({\bf M\to F})$-approach
 the number of fields used to describe membrane is reduced from $D$ to $D-n$.
  This reduction eliminates $n$ constraints of $({\bf V\to M})$-theories.
  $({\bf M\to F})$-approach action has no constraints.
  It allows us to find Hamiltonian formulation of theory
 in straightforward way in section \ref{Ham-sec}.

  $({\bf M\to F})$-approach is natural to represent continious
 distribution of membranes.
  Representation of Nambu-Goto type action in terms of set of scalar
 fields was initially suggested in papers \cite{hos1,hos2,hos3}.
  In \cite{morris1,morris2,morris3} the action was adopted to description
 of single membrane by insertion of delta-function.
  Invariance of equations of motion under coordinate transformation
 on ${\bf F}$ was demonstrated in papers 
 \cite{bf1,bf2}\footnote{The author is grateful to D. Fairlie,
  who brings his attention to references \cite{hos1}--\cite{bf2}.}.
  \cite{bf1,bf2} introduced alternative interpretation of the same
 $({\bf M\to F})$-action. 
  The action presented in the current paper is invariant under 
 coordinate transformations on both ${\bf M}$ and ${\bf F}$.
  Equations of motion for free delocolized membrane follow from
 energy-momentum conservation, meanwhile the form of energy-momentum
 tensor is determined by projective property.
  A generalization of p-brane solution is presented as an example of 
 delocalized membrane interacting with closed $(n+1)$-form.
  
\section{Notation}

  $D$-dimensional space time, i.e. smooth (pseudo)Riemannian
 manifold ${\bf M}$ with metric 
 $ds^2=g_{MN}dX^MdX^N$, $M,N,\dots=0,\dots,D-1$
 is considered.
  Small latin indices are used for space coordinates $m,n,\dots=1,\dots,D-1$.
  In further calculations ${\bf M}$ is considered to be
 a finite region in $\RR^D$ with smooth boundary $\partial{\bf M}$.

  For differential forms $A$ and $B$ with components
 $A_{M_1\dots M_q}$ and $B_{N_1\dots N_p}$
 the following tensor is introduced
$$
 (A,B)_{M_{k+1}\dots M_q~N_{k+1}\dots N_p}^{(k)}=\frac{1}{k!}~
         g^{M_1N_1}\dots g^{M_kN_k}
         A_{M_1\dots M_q}B_{N_1\dots N_p}.
$$
  Index $(k)$ indicates the number of indices to contract.
  If it can not lead to ambiguity, $(k)$ is skipped.

  For differential form of power $q$ it is
 convenient to introduce a norm $\|A\|^2=(A,A)^{(q)}$
 and Hodge duality operation $*A=(\Omega,A)^{(q)}$,
 where $\Omega=\sq~d^DX$ is form of volume.
  Here and below $g=\det(g_{MN})$.

  For set of differential forms $A^k$,
 numerated by index $k=1,\dots,m$
 the object $A^{\wedge n}$ is defined by the following
$$
  \left(A^{\wedge n}\right)^{k_1\dots k_n}
 =A^{k_1}\wedge\dots\wedge A^{k_n}.
$$
  Objects defined on manifold other then ${\rm M}$ bear
 additional index, the name of manifold.

\section{Projective properties of energy-momentum tensor}

  As a key property of membrane to reproduce, one has to
 choose a property, which can be written in terms of
 $D$-dimensional objects and has a simple geometrical meaning.
  It is natural to consider as the key property the
 following form of energy-momentum tensor $T_{MN}$
\be
  T_{MN}=-\rho~{\cal P}_{MN},
\label{TLP}
\ee
 where $\rho$ is tension in the direction specified by
 $n$-dimensional (${\cal P}^M_{~~M}=n$) orthogonal
 projector ${\cal P}^M_{~~N}$ (or {\em tension}).
  We refer the property (\ref{TLP}) as
 {\em projective property} of energy-momentum tensor.

\section{Delocalization of membrane action}

  The standard action for free membrane is defined as
 the world surface area.
  Integral is taken over the membrane world surface, i.e.
 along a manifold ${\bf V}$ of dimension $0<n<D$,
 meanwhile membrane geometry is described by fields $x^M(\xi)$,
 which define a map $x:{\bf V\to M}$.

  Generalizing standard free membrane action it is possible to introduce
 an action for delocalized membrane, defined by fields
 $x^M(\phi,\xi)$, which define a map $x:{\bf F'\times V\to M}$,
 meanwhile the integral is taken along $D$-dimensional manifold
 ${\bf F'\times V}$
\be
  S_{\bf F'\times V\to M}
  =-\int\limits_{\bf F'}d^{D-n}\phi\sqrt{|g^{\bf F}|}
   \int\limits_{\bf V}d^n\xi\sg,
\label{actFxV2M}
\ee
 where $\gamma=\det(\gamma_{ij})$,
 $\gamma_{ij}=g_{MN}(x(\phi,\xi))~\partial_i x^M~\partial_j x^N$,
 $\partial_i=\frac{\partial}{\partial\xi^i}$,
 ${\bf F'}$ is (pseudo)Riemannian manifold with metric
 $g^{\bf F}_{MN}(\phi)$.

  Similar to the case of standard membrane action,
 energy-momentum tensor satisfies projective property
 (\ref{TLP}), where $-\rho$ is equal to Lagrangian bulk density.

  Equations of motion for action (\ref{actFxV2M}) are calculated
 by variation by $x^M(\phi,\xi)$ and coincide with free membrane
 equations of motion
\be
  \frac{\delta S_{\bf F'\times V\to M}}{\delta x^M}
  =\partial_i\left(\sg~g_{MN}~\gamma^{ij}~\partial_jx^N\right)=0,
\label{eom-X}
\ee
 $\gamma^{ij}$ is matrix inverse to $\gamma_{ij}$.

  There are $n$ constraints, like in the case of standard membrane
\be
  \frac{\delta S_{\bf F'\times V\to M}}{\delta x^M}\partial_ix^M=0.
\label{svqzi}
\ee

  So, in the case $x^M(\phi,\xi)=x^M(\xi)$
 the model with action (\ref{actFxV2M})
 reproduces standard free membrane with tension
 $T=\int_{\bf F'}d^{D-n}\phi\sqrt{|g^{\bf F}|}$.

  The fields $x^M(\phi,\xi)$, describing delocalized membrane,
 define a map of $D$-dimensional manifold ${\bf F'\times V}$
 to another $D$-dimensional manifold ${\bf M}$.
  The case of special interest is the case then the map $x$
 is a (local) diffeomorphism, i.e.
 $\frac{Dx}{D(\phi,\xi)}\not=0$.
  At that time fields $x^M$ in action (\ref{actFxV2M})
 specify some coordinate transformation on the manifold
 ${\bf M=F'\times V}$.
  Equation $x^M(\phi,\xi)=X^M$ can be solved with respect to
 $\phi$ and $\xi$, which become functions of coordinates $X^M$:
 $\phi^\alpha=\varphi^\alpha(X)$,
 $\xi^i=\xi^i(X)$.
  Letters from the beginning of Greek alphabet numerate
 coordinates $\phi^\alpha$ on the manifold ${\bf F}$.
  Action (\ref{actFxV2M}) can be rewritten
 in terms of fields $\varphi^\alpha(X)$ and $\xi^i(X)$
 as integral over coordinates $X^M$.
  Action written in this form (see below) does not involve
 fields $\xi^i(X)$, i.e. it solves the constraints (\ref{svqzi}).
 
  Field $\varphi:{\bf M\to F\supset F'}=\varphi({\bf M})$
 with components $\varphi^\alpha(x^M(\phi,\xi))=\phi^\alpha$
 is referred as {\em membrane field}, its components $\varphi^\alpha$
 are referred as {\em membrane potentials}.
  Here ${\bf F}$ is some $(D-n)$-dimensional
 (pseudo)Riemannian manifold with metric $g^{\bf F}_{MN}(\phi)$.

\section{Membrane field}

  {\em Membrane field intensity} is defined as differential
 form of power $D-n$ of the following form
\be
  J=\left((d\varphi)^{\wedge(D-n)},\Omega_{\bf F}\right)^{(D-n)}_{\bf F}
  =\sqrt{|g^{\bf F}|}~d\varphi^1\wedge\dots\wedge d\varphi^{D-n}.
\label{def-J}
\ee

  $g^{\bf F}$ is a function of $\varphi$,
 so the form $J$ is closed.
  Under coordinate transformations on ${\bf F}$
 $g^{\bf F}$ transforms like determinant of
 (pseudo)metric on ${\bf F}$.
  Identity $dJ=0$ is natural to refer as
 {\em kinematical conservation law}.

  In further calculations we need also
 {\em unit normal}
 ${\rm n}=\frac{J}{\|J\|}$,
 and {\em unit tangent}
 $u=*{\rm n}$.

  Action is defined by formula
\be
  S_{\bf M\to F}=-\int\limits_{\bf M}d^DX\sq~\|J\|.
\label{actM2F-J}
\ee

  Action (\ref{actM2F-J}) is invariant under general
 coordinate transformation at ${\bf M}$ and ${\bf F}$.
  The following condition is also required $\|J\|\not=0$.

  Action (\ref{actM2F-J}) (in the case $|g^{\bf F}|=1$)
 was introduces for $n=2$ in papers \cite{hos1,hos2,hos3} as action
 for string.
  In papers \cite{bf1,bf2} the same action for arbitrary $D$ and $n$
 was considered as action for $(D-n)$-dimensional membrane.

  By variation of action (\ref{actM2F-J})
 over membrane potential $\varphi^\alpha$ one finds
 equations of motion
\be
%  \frac1{\sq}\frac{\delta S_{\bf M\to F}}{\delta\varphi^\alpha}=
  \sqrt{|g^{\bf F}|}~
  \partial_{M_1}\varphi^1\dots
  \widehat{\partial_{M_\alpha}\varphi^\alpha}
  \dots\partial_{M_{D-n}}\varphi^{D-n}~
  \nabla_{M_\alpha}
    {\rm n}^{M_1\dots M_{D-n}}=0.
\label{eom-n}
\ee
  In formula (\ref{eom-n}) factor under hat is skipped.

  By variation of action (\ref{actM2F-J}) over metric in
 space ${\bf M}$ one can find energy-momentum tensor
\be
  T_{MN}=-\|J\|~{\cal P}_{MN},
\label{emt}
\ee
 where projector is defined as ${\cal P}_{MN}=g_{MN}-({\rm n,n})_{MN}$.
  So, energy-momentum tensor satisfies projective property (\ref{TLP}).
  $\Tr{\cal P}=n$, so action (\ref{actM2F-J}) has to be considered
 as action for $n$-dimensional membrane.
  In \cite{bf1,bf2} this action was considered as action for 
 $(D-n)$-dimensional membrane.

  Equations of motion (\ref{eom-n}) follow from energy-momentum
 conservation and can be written as $\delta{\rm n}=0$,
 where $\delta=*^{-1}d*$.

  In is useful to mention also the following interesting property
 of action (\ref{actM2F-J}), equations of motion do not depend
 upon metric on manifold ${\bf F}$, which can affect only set
 of acceptable fields $J$.

  {\Large Theorem.}{\sl If membrane field intensity $J$ is
 spacelike, i.e. $\|J\|^2>0$, field
 $\varphi(X)$ and metric $g_{MN}(X)$ are continuously
 differentiable two times, 
 then equations of motion (\ref{eom-n})
 are equivalent to equations of motion (\ref{eom-X}),
 which are specified at all submanifolds $\varphi=const$.}

  So, action (\ref{actFxV2M}) and action
 (\ref{actM2F-J}) have the same energy-momentum tensors and
 equivalent equations of motion.

  In all formulae above, finiteness of $\|J\|$ and $\|J\|^{-1}$,
 was suggested, nevertheless, because equations of motion
 of membrane field have the form $\delta{\rm n}=0$,
 one can define equations of motion at the points, where
 unit normal has a limit.

  Moreover, solutions of equations of motion are transformed
 to solutions under the replacement of field
 $\varphi^\alpha(X)$ by $\psi(\varphi^\alpha(X))$,
 where function $\psi$ is monotone and continuously
 differentiable two times.
  Nevertheless, ambiguity of equations of motion let
 us postulate the same property for non-monotone
 functions $\psi$.
  So, if in some region of ${\bf M}$ $J=0$,
 then, it is natural to believe, that there is no field
 in this region, and equations of motion are satisfied
 identically.

  One can conclude, the action (\ref{actFxV2M}) and
 action (\ref{actM2F-J}) are equivalent not only in the case
 then $\frac{Dx}{D(\phi,\xi)}\not=0$, but also in more
 general class of situations.
  If $|g^{\bf F}|=1$, then
 the standard membrane ${\bf V}$ of unit tension defined
 as ${\bf V}=\{X\in{\bf M}|f^\alpha(X)=0\}$
 ($df^\alpha(X)|_{X\in{\bf V}}\not=0$)
 can be written in terms of field $\varphi$ by the following way
\be
  \varphi^\alpha(X)=\theta(f^\alpha(X)).
\label{theta}
\ee
  Substituting fields (\ref{theta}) into action (\ref{actM2F-J})
 we find
 $$S=-\int d^DX\sq \delta(f)\|df^1\wedge\dots\wedge df^{D-n}\|.$$
  This action was introduces in papers \cite{morris1,morris2,morris3}
 as action for regular (non-delocalized) membrane.

\section{Hamiltonian formulation\label{Ham-sec}}

  The action (\ref{actM2F-J}) has no constraints.
  It allows us to find Hamiltonian formulation of theory
 in straightforward way.

  Instead of $(D-n)$-form $J$ one can introduce
 $(D-n)\times(D-n)$ matrix
$$
  {\cal G}^{\alpha\beta}=g^{MN}\varphi^\alpha_{,M}\varphi^\beta_{,N}.
$$
  Let ${\cal G}_{\alpha\beta}$ be a matrix inverse to ${\cal G}^{\alpha\beta}$,
 and ${\cal G}=\det({\cal G}_{\alpha\beta})$.

  The action (\ref{actM2F-J}) can be written in the
 following equivalent form
\be
  S_{\bf M\to F}=
   -\int\limits_{\bf M}d^DX\sq~
    \sqrt{\det(g^{MN}\varphi^\alpha_{,M}\varphi^\beta_{,N})}=
   -\int\limits_{\bf M}d^DX\sqrt{\frac{|g|}{\cal G}}.
\ee
  Equations of motion (\ref{eom-n}) and energy-momentum tensor (\ref{emt})
 in terms of ${\cal G}$ acquire the form
\bea
  &&\partial_M
  \left(
    \sqrt{\frac{|g|}{\cal G}}~
    {\cal G}_{\alpha\beta}g^{MN}~\varphi^\beta_{,N}
  \right)=0,
\label{eom-G}
\\
  &&T_{MN}=-\frac1{\sqrt{\cal G}}
   \left(
      g_{MN}-{\cal G}_{\alpha\beta}\varphi^\alpha_{,M}\varphi^\beta_{,N}
   \right).
\label{emt-G}
\eea

  Momenta conjugated to fields $\varphi^\alpha$ are
\be
  p_\alpha=-\sqrt{\frac{|g|}{\cal G}}~{\cal G}_{\alpha\beta}~
  \partial^0\varphi^\beta.
\ee
  In terms of momenta 
\be
  {\cal G}^{\alpha\beta}=
\frac{\frac{{\cal G}^0}{|g|g^{00}}~
       h^{\alpha\gamma}p_\gamma~h^{\beta\delta}p_\delta}
     {1-\frac{{\cal G}^0}{|g|g^{00}}~
        h^{\kappa\lambda}p_\kappa p_\lambda}
  +h^{\alpha\beta}.
\ee
\be
  {\cal G}={\cal G}^0\left(1-\frac{{\cal G}^0}{|g|g^{00}}~
                h^{\alpha\beta}p_\alpha p_\beta\right).
\ee
  Here
\be
  h^{mn}=g^{mn}-\frac{g^{0m}g^{0n}}{g^{00}}
\ee
 is matrix inverse to space metric $h_{mn}=g_{mn}$,
\bea
  h^{\alpha\beta}&=&h^{mn}~\varphi^\alpha_{,m}~\varphi^\beta_{,n},
\\
  \frac1{{\cal G}^0}&=&\det(h^{\alpha\beta}).
\eea

  Using the formulae above, we write the Hamiltonian density
\be
  {\cal H}=\sqrt{\frac{|g|}{{\cal G}^0}
                 -\frac{h^{\alpha\beta}}{g^{00}}~p_\alpha p_\beta}
   -\frac{g^{0m}}{g^{00}}~\varphi^\alpha_{,m}p_\alpha.
\ee

  In Minkowski space-time, $g_{MN}=\diag(-1,+1,\dots,+1)$,
 Hamiltonian has the form
\be
  {\cal H}=\sqrt{\det(\varphi^\alpha_{,m}\varphi^\beta_{,m})
                  +\varphi^\alpha_{,m}\varphi^\beta_{,m}~p_\alpha p_\beta}.
\ee

\section{$p$-brane solution with delocalized source\label{p-b-sec}}
  As an example, let us consider solution of Einstein equations
 in the presence of field of antisymmetric tensor of power
 $q$ and membrane field interacting with the antisymmetric tensor.
  The solution presented generalizes $p$-brane solution in
 supergravity type model.

  Let us consider action
$$
  S=\int d^DX\sqrt{-g}
    \left(
      \frac{R}2-\frac{\|F\|^2}{2}
     -\|J\|+
%(-1)^{q(D-q)}
      \frac1h~(*J,A)
    \right),
$$
 where $F=dA$, $A$ is differential form of power $q$,
 $J=d\varphi^1\wedge\dots\wedge d\varphi^{D-q}$,
 $h=\sqrt{\frac{D-2}{q(D-q-2)}}$.

  By variation over fields $g_{MN}$, $A_{M_1\dots M_q}$, $\varphi^\alpha$
 one finds the following equations of motion
$$
  R_{MN}-\frac12Rg_{MN}=(F,F)_{MN}+\frac{(J,J)_{MN}}{\|J\|}
   -\left(\frac12\|F\|^2+\|J\|\right)g_{MN},
$$
$$
  \left(d\frac{*J}{\|J\|}+
%(-1)^{q(D-q)}
 \frac{F}h\right)\wedge (d\varphi)^{\wedge(D-q-1)}=0,
~~
  \delta F=(-1)^{q
%(D-q-1)
     }\frac{*J}h.
$$
  The fields defined by the following equations solve the
 equations of motion
$$
  ds^2=H^{-\frac2q}\eta_{ij}dX^idX^j
       +H^{\frac2{D-q-2}}\delta_{\alpha\beta}dX^\alpha dX^\beta,
$$
$$
  A=\frac{h}H~dX^0\wedge\dots\wedge dX^{q-1},
~~
  J=-h^2\triangle H~dX^q\wedge\dots\wedge dX^{D-1}.
$$
  Here $H$ is smooth positive function
 of space coordinates $X^\alpha$, $\alpha,\beta=q,\dots,D-1$,
 $\triangle=\sum_{\alpha=q}^{D-1}\partial_\alpha^2$,
 $i,j=0,\dots,q-1$, $\eta_{ij}=\diag(-1,+1,\dots,+1)$.
  Calculating $\|J\|$ one has to choose a branch of square root,
 which gives $\|J\|=-2h^2H^{-1-\frac2{D-q-2}}\triangle H$.

  In special case $D=4$, $q=1$ the solution above generalizes
 Madjumdar-Papapetru solution of Einstein-Maxwell equations
 for the case of the source, which is dust cloud with
 charge density equal to mass density.

  Similar to above generalization of $p$-brane
 solution, analogous generalizations of intersecting
 extremal $p$-brane solutions were found.
 (on intersecting $p$-brane solutions see
 \cite{AIR,IM} and references in these papers).
  These solutions are not presented in the article
 because of large volume of formulae, necessary to
 formulate the result in general case.

  The author is grateful to I.V. Volovich,
 Yu.N. Drozhzhinov and M.O. Katanaev.
  The work was partially supported by grant RFFI 99-01-00866.

%\newpage

\end{document}